\documentclass[pre,twocolumn,superscriptaddress,floatfix,showkeys]{revtex4}
\usepackage[dvips]{epsfig}
\usepackage{subfigure}
\usepackage{rotating}
\usepackage{endnotes}
\usepackage{amsmath}

\newcommand{\beq}{\begin{equation}}
\newcommand{\eeq}{\end{equation}}

\begin{document}
%
%

\title{Compaction dynamics in ductile granular media}
\date{\today}
%
%

\author{Lina \surname{Uri}}
\email{l.l.uri@fys.uio.no}
\affiliation{PGP, Department of Physics, University of Oslo}

\author{Dag Kristian \surname{Dysthe}}
\affiliation{PGP, Department of Physics, University of Oslo}

\author{Jens \surname{Feder}}
\affiliation{PGP, Department of Physics, University of Oslo}
%
%

\begin{abstract}
Ductile compaction is common in many natural systems, but 
the temporal evolution of such systems is rarely studied. 
We observe surprising oscillations in the weight measured 
at the bottom of a self-compacting ensemble of ductile grains.
The oscillations develop during the first ten hours of the 
experiment, and usually persist through the length of an
experiment (one week).
The weight oscillations are connected to the grain--wall
contacts, and are directly correlated with the observed 
strain evolution and
the dynamics of grain--wall contacts during the compaction.
Here, we present the experimental results and characteristic 
time constants of the system, and discuss possible reasons for the 
measured weight oscillations.
\end{abstract}


\maketitle

%
%
\section{\label{sec:intro}Introduction}
The stress distribution in dry granular media have been studied for 
more than a century. The German engineer Janssen studied the apparent
weight at the bottom of a silo as function of its filling 
height \cite{pap:Janssen1895}. 
Janssen found that the pressure at the bottom of a container of 
granular material increases linearly with small filling heights, 
but approaches a constant level exponentially slowly 
for large filling heights. 
That the measured weight at the bottom is less than the 
total weight of grains is referred to as a screening effect.
It is well known that the screening effect is due to the 
grain--wall friction 
 and how the stress distributes in a granular ensemble \cite{book:Duran99}. 
Janssen's mathematical expression for this, the Janssen law, compares
surprisingly well to experiments \cite{pap:Arroyo-Cetto03,pap:Vanel00},
in spite of its crude assumptions regarding friction and 
stress distribution \cite{pap:deGennes99}.
Over the last decade, various aspects of the stress distribution in 
static granular media have been studied. 
Experiments have shown that the stress distribution
is sensitive to the deposition history \cite{pap:Vanel99}, the shape
and size distribution of grains \cite{pap:Geng01a}, elastic properties
of the base \cite{pap:Brockbank97} and grains \cite{pap:Erikson02},
and that an exponential size distribution of forces is found at 
the bottom of a container for forces larger than the average 
\cite{pap:Liu95,pap:Lovoll99}.

The stress distribution in dynamic systems has been investigated
 in pushed columns of granular media inside a 
cylinder \cite{pap:Arroyo-Cetto03,pap:Bertho03,pap:Ovarlez03a,
pap:Ovarlez03b} 
by measuring the normal force at the bottom for constant driving velocities.
At small velocities, the measured force has a stick--slip 
behavior \cite{pap:Nasuno98,pap:Ovarlez03b} that is related
to aging of the grain--wall friction due to  
capillary condensation and shear strengthening of the contacts at the 
walls \cite{pap:Ovarlez03b}.
These dynamic systems consist of elastic particles, and the 
time dependence studied relate to other properties than the particle 
rheology.
In Nature, and in many technological processes, slowly compacted 
or sheared systems are dominated by the deformation of particles.
The time dependence in these systems is mainly given by the plastic
properties of the grains.

Here, the results from experiments on granular media
consisting of plastically deforming grains in a cylinder are presented. 
This system deformed slowly under its own weight, compacting 10\% 
in a week, while the normal force at the bottom (the {\it apparent mass} 
\cite{misc:ma}) was measured.
The initial expectation was that the system would show a granular 
Janssen type stress
distribution in the initial stage, but that due 
to the viscous rheology of the grains a stress distribution close
to the hydrostatic would develop.
Thus, the apparent mass was expected to increase.
Instead, the apparent mass developed unexpected (non-harmonic) oscillations, 
resembling the stick--slip behavior observed in hard granular 
media \cite{pap:Nasuno98,pap:Ovarlez03b}, except that it decreased 
initially, and the time scale of a ``slip'' could be up to an hour.
No overall increase was observed in the apparent mass after a week 
of compaction.
The strain development and wall contact dynamics were also studied during
the compaction, both showing behavior related to the weight oscillations.
The wall interaction between grains and cylinder was varied significantly in 
a few experiments, and proved crucial to the oscillations, as these
disappeared when wall friction was increased or decreased.
The experiments and results are described in the following two sections,
while some proposed mechanisms for the oscillations are discussed 
in section \ref{sec:discussion}.

We propose that the observed oscillations are due to two competing effects:
Grain--wall interaction opposing motion 
and the slow flow and relaxation of the 
grains inducing motion. 

\section{\label{sec:exp}Experiments}
\begin{figure}[floatfix]
\epsfig{file=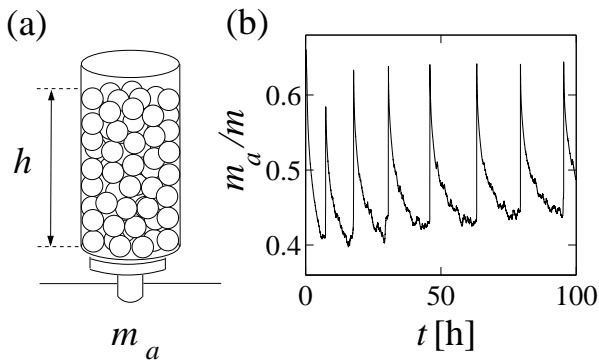,width=8.3cm}
\caption{(a) Schematic illustration of the setup. Ductile grains
were filled to a height $h$ in a cylinder and left to compact while
the apparent mass $m_a$ at the bottom was measured. (b) A typical 
recording of the apparent mass (shown normalized by the total mass
$m$ of the grains) as a function of time.}
\label{fig:setupres}
\end{figure}

We performed 30 experiments in which an ensemble of $N$ deformable grains 
were left to compact in a Plexiglas cylinder of diameter $D$.
The system was studied in several ways, but mainly by measuring the 
apparent mass $m_a$ at the bottom of the cylinder in order to 
follow the overall evolution of the stress distribution in the 
compacting ensemble.
A Mettler PM4800 balance was used to measure the apparent mass.
This balance operates by an induction mechanism that keeps the 
vertical position of the measurement area constant \cite{misc:Mettler}, 
and thus does not affect the compaction procedure.
The weight was measured 
to a precision of 0.03 g, and was typically a fraction (2--3)$\cdot 10^{-4}$
of the total mass of the grains.
The cylindrical container was mounted outside of the measurement
area.
Spherical grains were prepared manually from Play-Doh (Hasbro International 
Inc., UK) to a diameter $d=(8.8\pm 0.2)$ mm, and poured into the cylinder
approximately ten at a time.
The initial packing fractions were in the range $c=0.5$--0.6.
The material is viscous \cite{misc:visc} over 
a large range of strain rates, $\dot{\epsilon}=(10^{-2}$--$10^{-6})$ 
s$^{-1}$, with a viscosity of $\mu=3\cdot 10^5$ Pa\,s.
A schematic illustration of the setup is shown in Fig. \ref{fig:setupres}(a)
along with the typical result of the observed weight as a function of time.
The measured apparent mass $m_a$ presented in Fig. \ref{fig:setupres}(b) 
has been normalized by the total mass $m$ of the grains in the cylinder.
The apparent mass was found to oscillate in a quasi periodic manner. 
The period depended on details of the packing, and could increase or 
decrease slightly over the duration of each experiment. 
The filling height $h(0)$ at $t=0$ was varied between 1--4 times the 
cylinder diameter, and the cylinder diameter was varied between 3.4 and 
15 times the grain diameter.

\begin{figure}[floatfix]
\epsfig{file=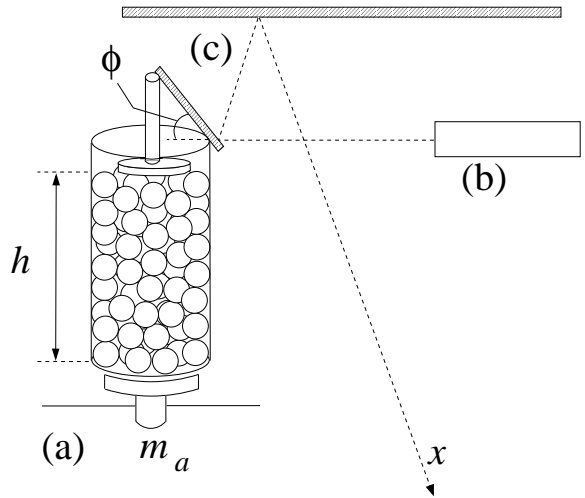,width=8.3cm}
\caption{Illustration of the experimental setup for strain
measurement by the use of mirrors and laser;  A balance (a) recorded the 
apparent mass $m_a$ at the bottom of the cylinder. 
The height of the packing
was measured as a function of time by a laser (b) beam that was reflected 
in a small and a large mirror (c), onto a point on the floor $x$. 
The position $x$ moved to the left as the angle $\phi$ between the 
small mirror and the horizontal plane was reduced, following the
reduction of the height $h$ of the compacting grains.
The piston rested on grains that did not touch the walls, thus the
strain was measured along a central axis.}
\label{fig:strain}
\end{figure}

In two experiments the total height $h(t)$ of compacting (granular) 
ensembles 
were measured using two different setups:
A camera was used in one experiment to take pictures of the compaction
process at various times.
Image analysis was then used to extract the height of the ensemble
based on the position of the uppermost 6 grains, to a 
resolution of 46 $\mu$m.
In another experiment, the height in the middle of the granular column 
was recorded by the use of a laser and two mirrors.
A small, light weight piston was placed on top of the central grains 
and allowed to move only vertically.
A small mirror was hinged onto the top of the piston, its lower end 
resting on the upper cylinder rim.
As the grains compacted, the mirror was tilted, reducing its angle $\phi$ 
to the horizontal plane.
A laser beam was reflected in the small mirror, and again in 
another, larger, mirror so that the beam was visible as a point on the floor.
The position of this point was recorded manually with time, and 
the height of the granular ensemble calculated to a precision of 3 $\mu$m.
The piston was positioned along the central axis of the container, and
followed the motion of the internal grains that did not touch the wall.
Figure \ref{fig:strain} illustrates the second strain measurement
method.
From the measurements of the total height the global strain, $\varepsilon$,
was found as $\varepsilon=1-h(t)/h(0)$.

The dynamics of the grain contacts at the cylinder wall was studied
using a camera (AstroCam, Capella, LSR Life Science Resources, 
UK) in one experiment.
The camera had a spatial resolution of 2000$\times$3000 square pixels, 
and 14 bit intensity resolution.
The contrast between the intensity inside and outside of a contact area
was within an 8 bit subset of the 14 bit dynamic range. 
The rim of a contact was established within two pixels with the 
spatial and intensity resolutions as specified.
The uncertainty that one extra rim of pixels introduced to the area
of a contact could be as high as 20\% for the smallest contact areas.  
The precision of the center of mass position was, however, much better,
as it does not depend on the exact choice of thresholding for the 
contact area.

\begin{figure}[floatfix]
\epsfig{file=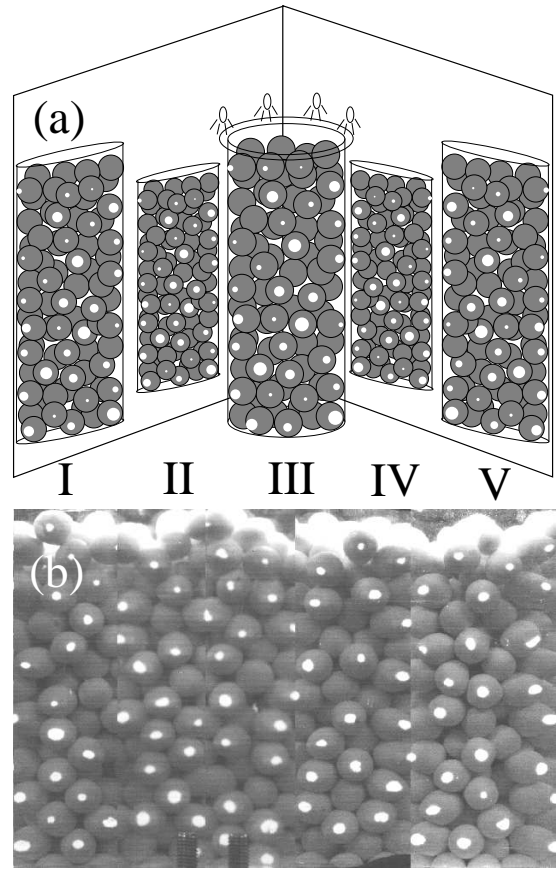,width=8.3cm}
\caption{(a) Schematic drawing of the setup for the measurement of 
contact areas at the wall of the cylinder. Two mirrors in an angle
$72^\circ$ to each other reflect the cylinder surface and the total area
can be extracted. Light emitting diodes were fitted into the top of the 
cylinder wall to enhance the contrast between contact regions (white) 
and regions of no contact (gray). 
(b) The unwrapped \cite{misc:Unwrap} surface after 
image treatment. Each of the five (I--V) cylinder images is scaled and 
unwrapped before they are fitted in overlapping regions. The match is 
only at the cylinder surface, which is 
why the internal regions seems mismatched in some places.}
\label{fig:mirror}
\end{figure}

The cylinder containing the ductile ensemble was placed in front of 
two mirrors which were set in an angle of $72^\circ$ to each other.
The cylinder was reflected twice in each mirror, thus the camera view
held five versions of the cylinder (I--V), capturing it from all sides. 
The grains' contacts to the wall were literally highlighted by shining
light down the Plexiglas wall of the cylinder. 
The light only reflected out of the wall in areas where the difference
in refraction indices were smaller than that between Plexiglas and air,
thus the contacts between grains and wall were bright in contrast to the 
pore space.
Figure \ref{fig:mirror}(a) illustrates the setup.
Each of the five cylinder images I-V (see Fig. \ref{fig:mirror}) 
was then `unwrapped' \cite{misc:Unwrap} and scaled according
to the geometry of the setup, then put together to form a continuous 
image of the surface area of the cylinder. 
An example of the resultant image is shown in Fig. \ref{fig:mirror}(b).
The spatial resolution in these images were 160 $\mu$m.
Images were recorded every 10 or 20 minutes for two days in order to capture
several oscillations.

A total of 90 contacts were recovered, and 79 of these were used in the 
analysis.
The remaining 11 contacts were discarded because of some mismatch of 
their area across boundaries between cylinder images.

An increase of the contact area of 70\% was observed during the 60 
hours that images were recorded, 60\% during the first 20 hours 
of compaction, and 10\% in the time interval $t\in[20,60]$ h.
A contact diameter was defined as $2\sqrt{A/\pi}$ for each contact 
area $A$, and found as a function of time.
The average contact diameter, $d_c$, was found by first taking the
average value of each contact diameter over the series of time steps in 
$t\in [20,60]$, and then find the average of this set,
$d_c=2.66\pm 0.02$ mm. 

\section{\label{sec:results}Results}
The typical behavior of the apparent mass $m_a$ in an experiment
is as follows:
At time $t=0$ all grains have been poured into the cylinder.
The apparent mass increases slightly over a period of a 
few minutes, reaches its maximum (often a global maximum) 
and then starts to decrease.
Weight oscillations mostly initiate during this decrease.
When oscillations have developed, their minima decrease toward a 
global minimum of $m_a$, before they increase slowly toward
a plateau.
The plateau varies between experiments in the range 45\%--88\% of the 
total mass $m$ of the grains,
but are mostly in the range 60\%--80\%. 

\begin{figure}[floatfix]
\epsfig{file=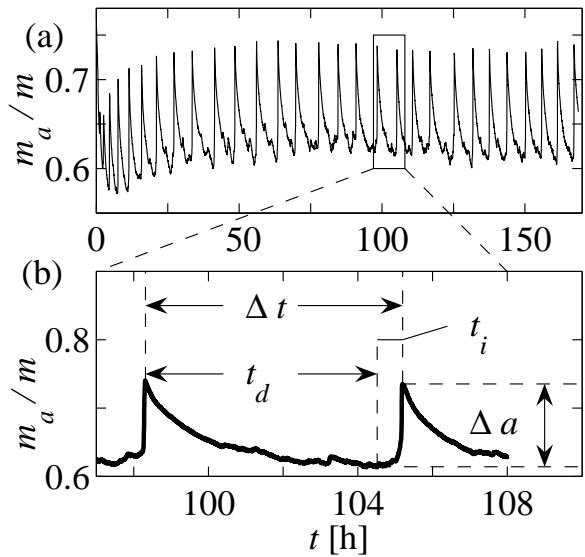,width=8.6cm}
\caption{(a) The evolution of the normalized apparent mass $m_a/m$ as
a function of time. (b) Closeup of one period. The total period $\Delta t$ 
is the time between two peaks. $t_d$ is the time through which the 
apparent mass decreases in one period, while $t_i$ is the time of 
increasing apparent mass. $\Delta a=\Delta m_a/m$ is the amplitude of an 
oscillation. A subscript $n$ is added to these parameters when they 
describe the specific values in oscillation number $n$.}
\label{fig:Pardef}
\end{figure}

Figure \ref{fig:Pardef} illustrates the definition of the periods, intervals,
and amplitude of an oscillation, which will be referred to in the
following.
The period $\Delta t$ of one oscillation is defined as the time between 
peaks $i$ and ($i+1$).
This period can be further divided into intervals $t_d$ and $t_i$ of 
overall decrease and increase, respectively, of the apparent mass. 
The point of minimum apparent mass between peaks $i$ and ($i+1$) 
marks the transition between the regions $t_d$ and $t_i$, see 
Fig. \ref{fig:Pardef}(b).
The amplitude $\Delta a$ of one oscillation is the change in normalized 
apparent mass $m_a/m$ during $t_i$.

\begin{figure}[floatfix]
\epsfig{file=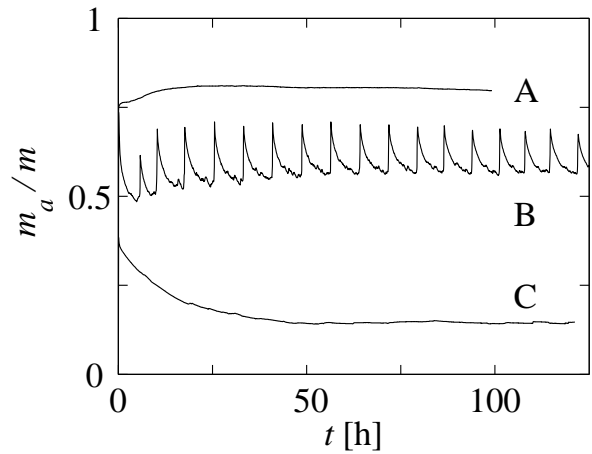,width=8.6cm}
\caption{The resulting apparent masses for different surface 
treatments:
Curve A was the result of coating the walls with Teflon (low friction). 
No coating of the Plexiglas wall resulted in curve B. 
Gluing sandpaper to the wall to enhance surface friction gave
curve C. }
\label{fig:Walls}
\end{figure}

The weight oscillations initially have small amplitudes, $\Delta a$, which 
increase toward a maximum after typically 3--16 oscillations.
The amplitudes reduce somewhat after this maximum value;
In some experiments they nearly disappear after 100 hours, while in 
others they are still at their maximum value after 200 hours.
The period $\Delta t$ of an oscillation also tends to increase 
initially, and then stabilize at a constant value after 
typically 17--80 hours.
In a few cases the period only stabilized after 150 hours, or not 
at all in the time span of the particular experiment. 
During $t_d$, irregularities larger than the typical noise level occur in 
$m_a/m$ in most of the experiments, see Fig. \ref{fig:Walls}, curve B.
These irregularities are referred to as ``micro-slips'' in the following.
Technically, a micro-slip, $dm_a^+$, is defined as the increase of $m_a/m$ 
in time intervals where the time derivative of $m_a$ is positive.

The observed oscillations in the apparent mass measured under the 
ductile granular ensemble was seen for all cylinder diameters and 
filling heights, and proved very robust 
to most perturbations applied to the system.
Varying the cylinder diameter and the filling height of grains did
not affect the amplitudes and periods in any consistent manner.
Amplitudes spanned 3\%--24\% of the total mass and the periods were 
in the range $\Delta t=(0.7$--47) h when all experiments are 
considered.
Two otherwise equal experiments could produce different characteristics,
in one case producing amplitudes of 6\% and 20\%, and periods of 
3.8 and 7.3 hours, respectively.
The variability is probably due to details of the random packings that is 
beyond experimental control.
 
Changing the surface properties on the cylinder wall was the only 
perturbation that dramatically affected the oscillations.
Figure \ref{fig:Walls} shows results from experiments in which 
the surface friction was reduced by Teflon (curve A), and enhanced
by (400 grit) sandpaper (curve C). 
In the following these experiments are referred to as `the Teflon-'
and `the sandpaper experiment', respectively.
No alteration was done to the surface of the wall in the experiment 
that produced curve B, which, apart from the surface, was identical to 
the Teflon- and sandpaper experiments.
As can be seen from the figure, reducing or enhancing the wall
friction both removed the weight oscillations.
By reducing the friction on the wall the apparent mass increased
slightly from the initial value (curve A, Fig. \ref{fig:Walls}).
Although Teflon reduced friction considerably, it did not remove
it fully, which would have made the apparent mass equal to the
total mass of the grains.

\begin{figure}[floatfix]
\epsfig{file=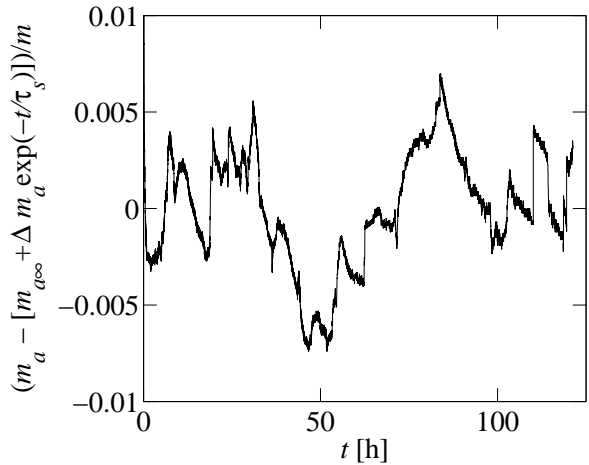,width=8.6cm}
\caption{Deviations of the fit from the measured normalized apparent mass,
$m_a/m$, as a function of time $t$ for the sandpaper experiment, see
Eq. \ref{eq:FitSandpaper}.}
\label{fig:DevFitSandpaper}
\end{figure}

By increasing wall friction another behavior emerged, as the 
apparent mass decreased, apparently toward
a constant level (curve C, Fig. \ref{fig:Walls}).
Curve C was fitted excellently by 
\beq
m_a/m=(m_{a\infty}+\Delta m_a \exp{[-t/\tau_s]})/m\,,
\label{eq:FitSandpaper}
\eeq 
where $m_{a\infty}=(7.027\pm 0.001)$ g, $\Delta m_a=(10.830\pm 0.005)$ g,
and $\tau_s$ is a characteristic time constant of $(13.52\pm 0.01)$ h.
The uncertainties are the likely error in the best fit parameters.
Figure \ref{fig:DevFitSandpaper} shows the deviations between the data and 
the fit, $(m_a - m_{a\infty}+\Delta m_a \exp{[-t/\tau_s]})/m$.
The exponential decay fits the observation exceptionally well, and the 
deviations are within the range $[-0.0077, 0.0076]$ of the normalized data.
Nevertheless, micro-slips are easily recognizable above the experimental
noise, which is of the order of $2\cdot10^{-4}$ (0.03 g/142.6 g) of 
the normalized apparent mass, while the slips are of the order 
of $7\cdot10^{-3}$.
The experimental noise is not visible in the figure.

\begin{figure}[floatfix]
\epsfig{file=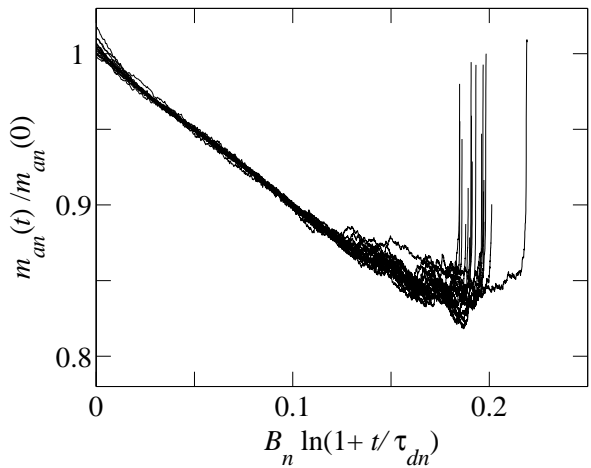,width=8.6cm}
\caption{The decreasing apparent mass of 17 oscillations in one experiment
(see Fig. \ref{fig:Pardef}(a))
plotted as a function of time according to Eq. \ref{eq:FitOsc}. 
(The expression on the horizontal
axis is the time dependent part of Eq. \ref{eq:FitOsc}.)}
\label{fig:FitOsc}
\end{figure}

A fit has also been made to the decreasing part $t_d$ of each oscillation 
$n$ in the curve of Fig. \ref{fig:Pardef}(a), which is consistent with 
logarithmic decay with time:
\beq
\frac{m_{an}(t)}{m_{an}(0)}=\big(1-B_n\ln{[1+t/\tau_{dn}]}\big)\, .
\label{eq:FitOsc}
\eeq
Here, $m_{an}(t)$ is the apparent mass of the $n$-th oscillation, and
$m_{an}(0)$, $B_n$ and $\tau_{dn}$ are best fit parameters to the equation,
calculated for each oscillation $n$ separately. 
$m_{an}(0)$ is the best fit value of $m_{an}$ at the start of the decrease, 
based on the first 2.5 h of the decreasing $m_{an}$.
$\overline{m_{an}(0)}=76.4\,[-0.4,0.5]$ g is the median value of $m_{an}(0)$,
 with the quartile deviations in brackets. 
$\overline{B_n}=0.042\,[-0.002,0.004]$ is the median of the set of 
dimensionless constants $B_n$, and 
$\overline{\tau_{dn}}=0.16\,[-0.02, 0.03]$ h is the median and quartiles
of the set of $\tau_{dn}$.
Figure \ref{fig:FitOsc} shows the collapse of the weight data when 
plotted according to Eq. \ref{eq:FitOsc}.
The limited dynamic range on both axes suggests that one can also fit 
the data by a power law with a small exponent. 
We have not found any theoretical arguments for the choice of one fit over
the other, thus the main observation is that the decreasing parts of the 
oscillations have the same form over the first 2.5 hours,
with a time constant of $\overline{\tau_{dn}}=0.16$ h.
The sandpaper gave a decreasing exponential function with time, as 
gives the initial decrease of $m_a$ during $t_d$ in an oscillation:
\beq
\lim_{t\rightarrow0}1-B\ln{(1+t/\tau_d)}\simeq 1-Bt/\tau_d\simeq
\exp{(-Bt/\tau_d)}=\exp{(-t/\tau_0}\,.
\eeq
The functional dependence is thus similar to the sandpaper at the 
start of the decrease, with a time constant of $\tau_0=\tau_d/B=3.8$ h.

\begin{figure}[floatfix]
\epsfig{file=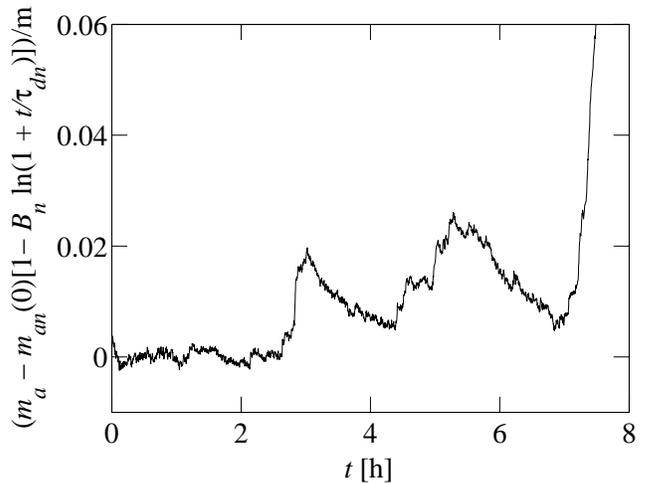,width=8.6cm}
\caption{The deviations from the measured $m_a/m$ of its fit for the 
decreasing part of one oscillation, as a function of time, see 
Eq. \ref{eq:FitOsc}. }
\label{fig:DevOsc}
\end{figure}

The deviation from the fit is plotted for one oscillation in Fig. 
\ref{fig:DevOsc}.
Large deviations on the order of 2\% of the total mass (the micro-slips) 
develop some time into $t_d$ (typically 3 hours in this experiment).  
All visible irregularities in this plot is above the noise level of 
the measurements.

\begin{figure}[floatfix]
\epsfig{file=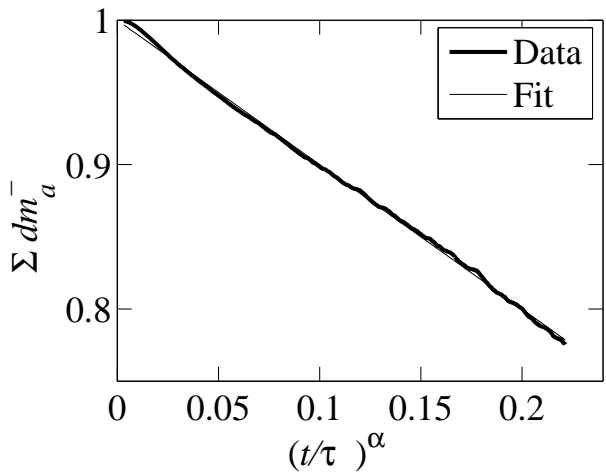,width=8.6cm}
\caption{The cumulative sum of decreasing $m_a$ during an oscillation
as a function of the scaled time $(t/\tau_-)^\alpha$, see Eq. 
\ref{eq:Fitminus}.}
\label{fig:FitMinus}
\end{figure}

Taking the time derivative of $m_a$ as 
$dm_a/(m\,dt)=(m_a(i+1)-m_a(i))/{m[t(i+1)-t(i)]}$, the set of positive 
increments of $m_a/m$ (the micro-slips, $dm_a^+$) and negative increments ($dm_a^-$) were
found for each oscillation's $t_d$.
The micro-slips were removed from the decreasing part of the oscillations
by cumulative summation of $dm_a^-$, and the resulting data set fitted
by a power law,
\beq
\sum_n dm_a^-(t) -1=-(t/\tau_-)^\alpha\,.
\label{eq:Fitminus}
\eeq
The median and quartiles of the fitting parameters are $\alpha=0.55$ 
[-0.02, 0.02], and $\tau_-=130$ [-37, 24] h, see Fig. \ref{fig:FitMinus}.
No characteristic time exists for a power law, since 
$-\lambda^\alpha(t/\lambda\tau)^\alpha$ fits equally well for all $\lambda$.

\begin{figure}[floatfix]
\epsfig{file=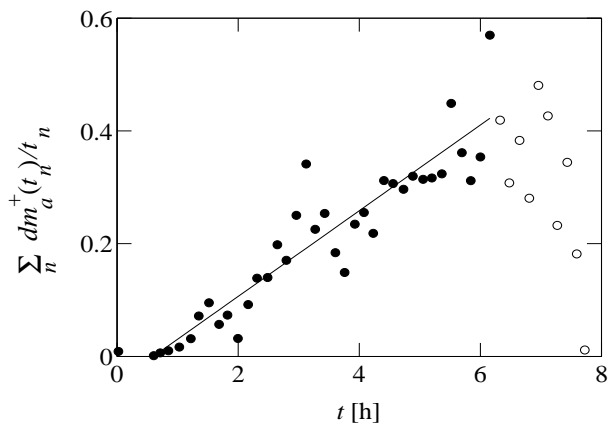,width=8.6cm}
\caption{The micro-slip `activity' as a function of time after each 
big slip. The activity is found as the sum of micro-slips, $dm_a^+/m$, 
from all oscillations in Fig. \ref{fig:Pardef}, binned in times $t_n$
and normalized by the width of the bin.}
\label{fig:maPlus}
\end{figure}

The micro-slips $dm_a^+$ were found as a function of time in all the 
oscillations of the experiment shown in Fig. \ref{fig:Pardef}, and 
binned in 50 time intervals.
The sum of micro-slips was taken for each bin, and divided by the size
of the bin to produce the temporal evolution of micro-slip `activity'.
Figure \ref{fig:maPlus} presents the result.
As the $t_d$ were of different lengths for each period, not all bins 
contain contributions from all oscillations.
The bullets present times that include data from all oscillations, 
whereas a circle includes only data from $t_d$  
long enough to contribute to the specific bin.
The line through the data is a linear fit, based on all but the first
bullet, given by $\sum_n m_a^+(t_n)/t_n =A(t-t_0)$.
Here, $A=(0.076\pm0.005)$ h$^{-2}$, and $t_0=(0.6\pm 0.2)$ h.  
The activity presented by the bullet at $t\sim0$ is probably 
remnants from the big slip that occurred at $t=0$, thus the micro-slip
activity is initiated at time $t_0$ after each big slip and grows
linearly until another big slip occurs.

We could not find a model with few parameters that would fit the 
`Teflon' results (curve A in Fig. \ref{fig:Walls}) due to the 
complex initial evolution.
This curve also shows some micro-slips, of size $1.5\cdot10^{-3}$, 
larger than the noise level of $4\cdot10^{-4}$. 

\begin{figure}[floatfix]
\epsfig{file=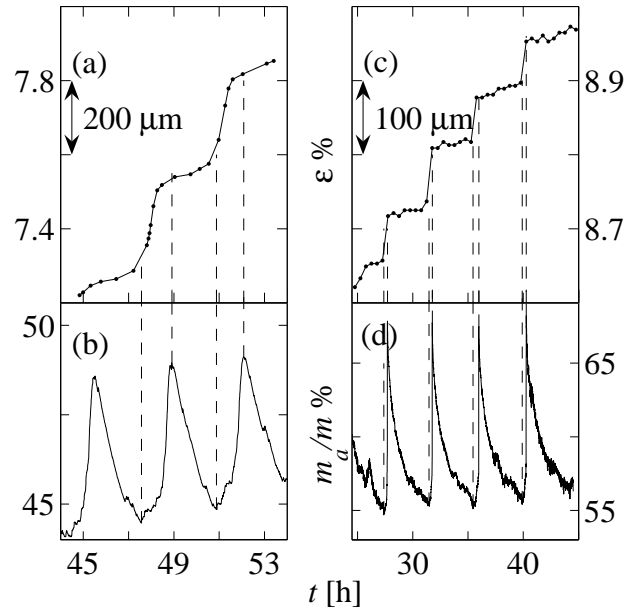,width=8.6cm}
\caption{Details of the strain as a function of time, measured in two 
experiments compared to the 
weight $m_a$. 
(a) The global strain, $\varepsilon$, measured with 3 $\mu$m
resolution (see Fig. \ref{fig:strain}) as a function of time. 
(b) The normalized apparent mass, $m_a/m$, as a function of time for 
the experiment in (a).
(c)$\varepsilon$ measured with 46 $\mu$m resolution by the high
resolution camera. 
(d) The apparent mass as a function of time for the experiment in (c).}
\label{fig:strains}
\end{figure}

The measurements of the height of the system as function of time
revealed that the vertical motion occurs in steps.
This was seen in both strain experiments, and is shown in 
Fig. \ref{fig:strains}(a) and (c).
Figures \ref{fig:strains}(b) and (d) show the simultaneous 
measurements of the normalized apparent mass, $m_a(t)/m$.
From the experiment with 3 $\mu$m resolution, the minimum and maximum
compaction velocities of the central part of the cylinder were found 
to be $5.4\cdot 10^{-9}$ m/s and $7\cdot 10^{-8}$ m/s, respectively.
The maximum acceleration, occurring at the start of a compaction step,
was $1\cdot 10^{-11}$ m/s$^2$.
Comparing the region of decreasing $m_a$ of Fig. \ref{fig:strains}(b)
to the strain in (a), a small but visible vertical movement  
occurs along the central axis of the packing during the weight decrease.
The main increase of strain during one oscillation (that is, the step)
takes place within the region in which the apparent mass increases 
from its minimum to its maximum.
Unfortunately, the limited resolution of the strain measurements in 
Fig. \ref{fig:strains}(c) prevented a detailed comparison
between the strain evolution of the 6 uppermost grains and the 
apparent mass.
It is evident from this measurement, however, that the global strain
motion is directly correlated with the changes in the apparent mass.
A compaction velocity of the uppermost grains of (0.6--3)$\cdot 10^{-9}$
m/s was found during $t_d$, and 4$\cdot 10^{-8}$ m/s during $t_i$.

The dynamics of the wall contacts were studied in one 
experiment as described in section \ref{sec:exp}.
Having found the `unwrapped', properly scaled surface of the cylinder
(see Fig. \ref{fig:mirror}), we obtained a high contrast of the contact.
The development of the 
area and center of mass position of each contact was followed through 
the experiment.

The weight oscillations correlated strongly to the contacts' center
 of mass motion, while no such correlation was found with the changes
in contact area.

\begin{figure}[t]
\epsfig{file=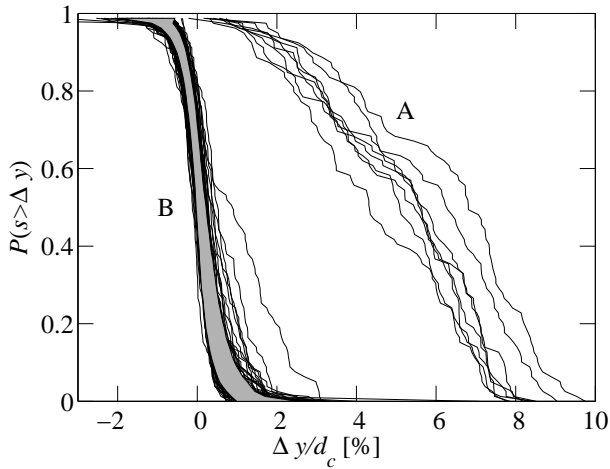,width=8.6cm}
\caption{The cumulative distribution $P(s>\Delta y)$ as a function of 
normalized vertical contact displacement $\Delta y/d_c$ between two
consecutive images. Motion 
downward has positive values of $\Delta y$.
Curves A result during slips, while curves B present the remaining 
movement between time steps. The gray region through B covers the average value
taken at each 1/79 interval of $P$, plus and minus
the standard deviation from the average.}
\label{fig:Ydisp}
\end{figure}

The contacts were seen to move ``simultaneously'', that is, within the
temporal resolution of the images, which means they all slipped 
within a period of 15--20 minutes, during $t_i$.
Figure \ref{fig:Ydisp} shows the cumulative distribution $P(s>\Delta y)$ 
of vertical contact displacement $\Delta y$ between two consecutive 
images.
The contact displacement is normalized by the average contact diameter
$d_c$. 
Each curve corresponds to the distribution in one time step of the experiment.
Curves A present the motion during slips, while curves B are the motion 
in time steps between slips.
The gray band through B spans the average $\pm$ the standard deviation
of vertical motion.

The median vertical displacement of a contact during a slip was 6 [$-1, 2$]\% 
of the average contact diameter, $d_c$.
Outside of the slips the median displacement was only 0.07 [$-0.20, 0.24$]\%
of $d_c$.
\begin{figure}[th]
\epsfig{file=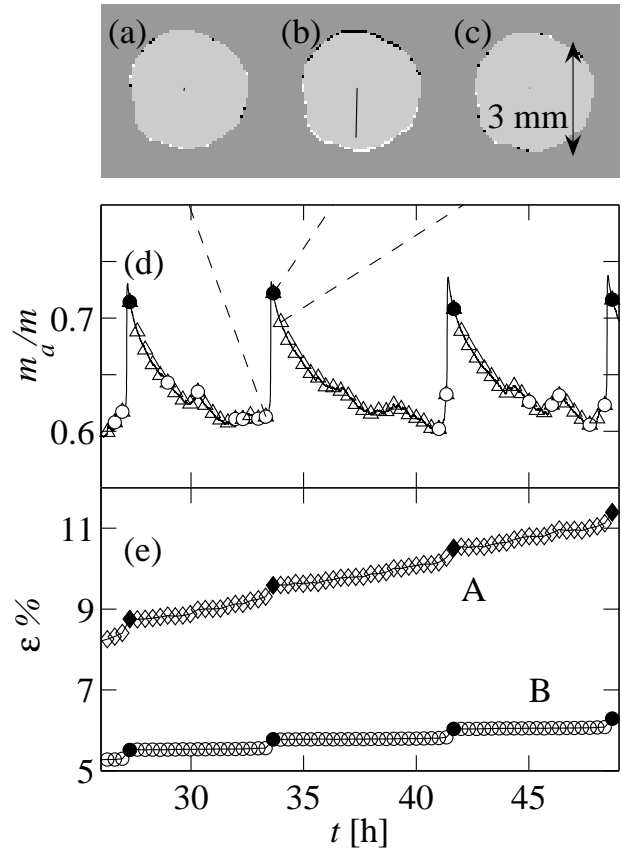,width=8.6cm}
\caption{(a), (b) and (c) shows difference images of a contact between
consecutive images. White is newly established area, black is area that
no longer is a part of the contact, and light gray is the unchanged contact
area. The center of mass motion between the images are shown as black 
lines. The length of the lines is exaggerated 10 times.
(d) shows the normalized apparent mass, the triangles
($\triangle$) mark the times when pictures were taken of the ensemble.
Circles ($\circ$) mark the times when minimum 15\% of the contacts 
moved more than 1\% of the average contact diameter. 
Bullets ($\bullet$) mark the 
times when more than 80\% of the contacts moved at least 2\% of the 
average contact diameter. 
The lower plot (e) shows the average strain development found from image 
analysis for the 20 lower
($\circ$, curve B) and upper ($\diamond$, curve A) wall contacts. 
Filled symbols represent the times that a picture was taken at or 
immediately after a peak in the apparent mass presented in (d).}
\label{fig:contacts}
\end{figure}
Figures \ref{fig:contacts}(a), (b) and (c) show the difference in one
contact area between consecutive images in one experiment. 
White corresponds to new contact area, black to area that was left since
the previous image, and light gray shows contact area where no changes
occurred.
Figures \ref{fig:contacts}(d) and (e) show the normalized apparent 
mass and the average strain of the upper (diamonds) and lower 
(circles) 20  wall contacts, respectively.
The markers in both plots represent the times when pictures were 
taken.
In Fig. \ref{fig:contacts}(d) circles mark the times when 15\% of the
contacts moved more than 1\% of the average contact diameter in 20 minutes
(since the last image). 
The bullets show the times when 80\% of the contacts moved at least
2\% of the average contact diameter. 
Triangles represent the times when pictures were taken.
Based on the observed area of the grain--wall contacts and the 
measured $m_a$, the average load per square millimeter carried by a 
contact was calculated to be in the range (0.5--1.2) kPa.

Table \ref{tab:param} presents the characteristic values of various 
parameters: (a) gives the median period, amplitude, intervals $t_d$
and $t_i$, and characteristic times $\overline{\tau_d}$ and $t_0$ 
for the oscillations in one experiment (see Fig. \ref{fig:Pardef}).
(b) presents the characteristic time from the fit of $m_a/m$ from the 
sandpaper experiment, and the estimated characteristic time of elastic 
relaxation (see section \ref{sec:discussion}).
\begin{table}
\begin{tabular*}{\linewidth}{@{\extracolsep{1cm minus 1cm}}ccl}
\hline
\hline
\multicolumn{3}{c}{Characteristic values}\\
\hline
&$\overline{\Delta t}$& 6.4 [$-$0.7, 1.2] h\\
&$\overline{t_d}$&5.2 [$-$0.3, 1.1] h\\
&$\overline{t_i}$&0.8 [$-$0.3, 1.0] h\\
\hspace{0.8cm}(a)\hspace{0.8cm}&$\overline{\tau_d}$&0.16 [$-$0.02, 0.03] h \\
&$\tau_0 = \overline{\tau_d}/B$& $\sim$ 3.8 h\\
&$t_0$&0.6 $\pm$ 0.2 h\\
&$\overline{\Delta a}$&12.6 [$-$0.3, 1.0] \%\\
\hline
&$\tau_s$&13.52 $\pm$ 0.01 h\hspace{5mm}\\
\raisebox{2mm}[0cm][0cm]{\hspace{0.8cm}(b)}&$\tau_e$\hspace{5mm}&%
$\sim 10^{-6}$ h\\
\hline
& $l_s$&$\sim 260\,\mu$m\\
& $l_0$&$\sim 74\,\mu$m\\
\raisebox{2mm}[0cm][0cm]{\hspace{0.8cm}(c)}&$l_d$&$\sim 101\,\mu$m\\
&$l_i$&$\sim 115$--$200\,\mu$m\\
\hline
\hline
\end{tabular*}
\caption{(a) Median values of the period $\Delta t$, amplitude $\Delta a$,
the intervals $t_d$ and $t_i$, the characteristic times 
$\overline{\tau_d}$ and $\tau_0=\overline{\tau_d}/B$ of decreasing $m_a$, 
and $t_0$ of activation of 
micro-slips of the experiment presented in Fig. \ref{fig:Pardef}(a).
(b) Characteristic times $\tau_s$ of the $m_a/m$
evolution in the sandpaper experiment, 
and $\tau_e$, the estimated time of relaxation of elastic stress.
(c) Estimated characteristic length scales, from time scales in (a) and (b),
see section \ref{sec:discussion}.}
\label{tab:param}
\end{table}

\begin{figure}[floatfix]
\epsfig{file=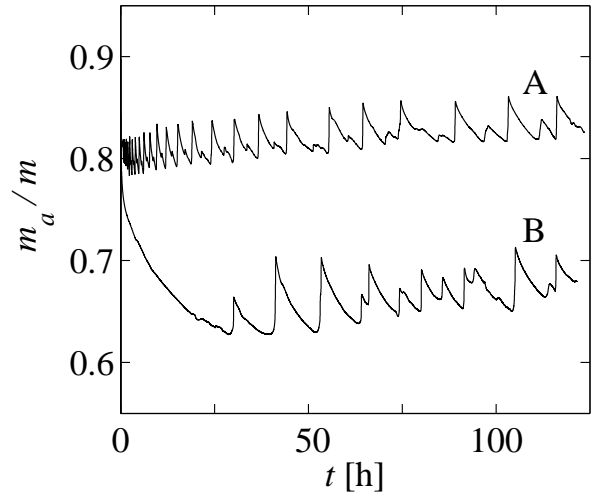,width=8.6cm}
\caption{The normalized apparent mass, $m_a/m$, as a function of time, $t$, 
at the bottom of a cylindrical 
Play-Doh sample (B), as 
compared to $m_a/m$ from a granular geometry (A) as a function of time. }
\label{fig:bulk}
\end{figure}

One experiment was performed to understand how the granular geometry of 
the ensemble affected the apparent mass. 
The granular ensemble was exchanged with a non-porous slab of Play-Doh
that did not fill the cylinder, but touched both 
the bottom and the walls of the setup. 
This experiment is referred to as `the bulk experiment' in the following.
The slab was left to flow into the available space, and the 
apparent mass was measured as before, see curve B of Fig. \ref{fig:bulk}.
Again, a granular version of this experiment was conducted for 
comparison, in which the total mass of the grains and the cylinder 
diameter were the same as those of the bulk experiment, see Fig. 
\ref{fig:bulk}, curve A.
As seen from the figure, both setups produced weight oscillations,
thus the granular geometry is not the (only) reason for the oscillations.
Oscillations started later in the bulk case than in the granular
case, and both systems show uncommon irregularities in their periods.
The granular system had nearly 100 grain--wall contacts, while 
the bulk sample had 3--4 large contact areas. 
The oscillations are probably due to the multi-contact nature of the 
interface between the deforming sample and the confining cylindrical 
wall.

\section{\label{sec:discussion} Discussion}

The self-compaction of a ductile ensemble 
depends on the deformability of the grains and on a porous structure.
The granular geometry of the ensemble was not necessary for 
oscillations to form, as weight oscillations also resulted 
under a bulk slab of material that deformed viscously into 
available space.
This result emphasizes the importance of the multi-contact wall
interaction to the observed oscillations in the apparent mass.

The grain--wall interaction proved to be crucial to the oscillations
in the apparent mass by the experiments with 
varying wall friction.
No oscillations were observed when increasing or decreasing the wall 
friction from that of the regular  experiments with Plexiglas walls.
The evolution of $m_a$ in these experiments is interesting because it 
shows two different behaviors according to the wall friction.
A low wall friction resulted in an increasing apparent mass, while 
a high wall friction made the measured weight decrease.
The same mechanisms leading to these results are likely to be
the reason for the oscillations observed in $m_a$ in the 
regular experiments.

The reason for the decrease of the apparent mass must be that the
walls sustain an increasing part of the total grain mass, that is, 
a dynamic vertical force must act on the grain contacts from the wall.
This force could be friction, shear cohesion, or a combination of 
the two, and will be referred to as the grain--wall interaction
in the following.

The increasing weight was initially believed to be due to a new
internal grain--grain contact.
As the stress distribution in hard granular media is known to be
very sensitive to local arrangements of grains, a new contact
was believed to change the stress distribution.
New contacts would preferentially form in the vertical direction,
because of the anisotropic compaction, and thus would tend to 
redirect stresses toward the bottom.
The number of new contacts in the ensemble is limited, and the
average number of contacts per grain increased from 6.5 to 
7 \cite{pap:Uri05a} during a typical experiment.
50--100 new contacts were expected to form during an experiment, 
which is roughly twice the typical number of oscillations.
If we assume that not all new contact formations are noticed in $m_a$,
perhaps because of their position in a low stressed region, 
that would explain the shortage of oscillations, and the 
micro-slips in $m_a$ in the oscillations and the sandpaper 
experiment.
On the other hand, this assumption directly disagrees with the 
nearly constant amplitudes seen in all experiments.
The experiment of a bulk slab, which also produced weight oscillations, 
eventually proved that new internal contacts between grains were not
the main reason for the weight oscillations.
Stress redistribution is, however, thought to take place continuously
during the slow internal flow of material, both in the granular and 
the bulk systems.

In principle, elastic energy could be stored in compressed parts
of the packing after a slip, resulting in a decreased grain--wall
interaction. 
The relaxation of this elastic energy could cause the observed
decrease in the apparent mass.
The characteristic time of elastic relaxation is expressed as
the ratio of viscosity $\mu$ to bulk modulus $K$.
We know that the viscosity of Play-Doh is of the order of $10^5$ Pa\,s
for shear rates as low as $10^{-6}$ s$^{-1}$.
The bulk modulus was not measured, but it is expected to be closer
to that typical of fluids ($K_f\simeq$ 1--2 GPa, \cite{book:scidata}) 
than that of iron ($K_i\simeq 170$ GPa, \cite{book:mathand}), thus 
on the order of $10^9$ Pa.
The resulting estimate of elastic relaxation time for Play-Doh is
\beq
\tau_e=\mu/K=10^{5}/10^9=10^{-4}\text{s}\, .
\eeq
Elastic compressive stresses should relax in (less than) seconds, 
which is much less than any time scale observed in an oscillation.

Another explanation for the decreasing $m_a$ emerges from 
the assumption that the ratio of horizontal to vertical stresses increases 
with increasing packing fraction.
If friction is assumed to be proportional to the normal force,
an increasing horizontal stress in the packing would result in 
increased wall friction, hence a decrease in $m_a$.
The packing fraction increases approximately 10\% during the experiment,
while the characteristics of the oscillations does not change.
This implies that the packing fraction is 
not the main parameter for describing the dynamic behavior.

The reason for a decreasing apparent mass can be seen in connection 
to the shearing of grain--wall contact regions.
During the time $t_d$ of decreasing $m_a$ the strain increases
very slowly, suggesting that only an internal flow of grains contributes to
the strain in this regime (see Fig. \ref{fig:strains}(a) and (b)).
The analysis of the motion of grain--wall contacts shows that 
the vertical motion of contacts in this regime is limited and noisy,
thus most contacts are practically at rest (even though the central 
part of the packing creeps).
Due to the slow flow internally in the packing, they are also 
continuously sheared.

There are clear slips of the order of 2\% in the normalized apparent 
mass during the 
decreasing part of the period in the granular setups (see Figs. 
\ref{fig:Pardef} and \ref{fig:DevOsc}). 
Micro-slips are not seen in the weight data from the bulk experiment,
thus their origin seems to be the granular geometry, or possibly the
large difference in the number of wall contacts between the bulk 
and granular systems.
No collective motion is seen at the wall during the micro-slips in the
granular experiment, although 5\% of the contacts move a distance of 1\%
of $d_c$ in every time step, thus their motion might be connected 
to the measured micro-slips in $m_a$.
Micro-slips might be due to the internal reorganization of forces 
within the granular system, which may trigger 
some of the grain--wall contacts into motion.
A reorganization of forces must also take place in the material in the
bulk experiment, although probably in a different way than 
that of the more complex granular geometry.
The reorganization must increase the average shear stress in the contact 
regions, which again leads to an increase of the vertical grain--wall
interaction.
Once a contact experiences a shear stress that can not be sustained by 
the grain--wall interaction, it ``breaks'', or starts to move.

The strain development could not be measured in the 
sandpaper experiment, thus whether this system compacted much is 
not known.
Similar, but smaller, micro-slips than those seen in regular experiments 
were seen in the sandpaper experiment.
This suggests that internal stress rearrangement was taking place. 
The grain--wall interaction was considerably higher in the 
sandpaper experiment than in the regular setup (as the apparent mass
reached a minimum of 15\% of the total mass).
It is reasonable to assume that the contacts did not move much, 
or in any correlated manner, based on the lack of weight oscillations.


The direct correspondence between the step in the strain and 
the increasing $m_a$ in Fig. \ref{fig:strains}(a) and (b) implies
that the motion of wall contacts is very important for the weight increase.
Assuming that wall contacts are broken, or mobilized, at a critical 
shear stress, one or more contacts will initiate a slip, and 
the others follow.
The contacts that break contribute to a decrease in the total 
wall interaction, thus a possible increase of the apparent mass.
The sum of wall interactions decrease over a time period
that must depend on how fast the contact breaking propagates among 
the other contacts, and how fast each contact breaks.
From our temporal resolution in the study of grain--wall contacts,
we see that all grains move within 20 minutes in connection
with the slip.
The apparent mass will increase according to the decreasing wall 
interaction $F_w$, as the force balance of the system is
$\sum F=m_a\,g -F_w = m\,a < 10^{-12}$ N $\simeq 0$.

The strain development was not measured in the
Teflon experiment, thus it 
is not known whether the strain had similar steps during the
compaction as in the regular experiments.
Based on the direct correlation between weight oscillations and 
the observed strain in the regular experiments, however, it seems likely 
that the wall contacts in the Teflon experiment in some sense moved 
continuously, as no oscillations in $m_a$ were observed here.
Micro-slips were observed in $m_a$, however, thus some dynamic
interaction between the grains and the wall was present, probably
because of internal stress rearrangements.
Sliding contacts also support some of the grain mass, as neither 
during $t_i$ in the regular experiments nor in the Teflon experiment 
does the apparent mass reach 100\% of the grain mass.
The grain--wall interactions during motion are smaller, however,
than in the static case, as the apparent mass increases during 
motion in the regular experiments, see Fig. \ref{fig:strains}.

That all contacts are mobilized within a time interval corresponding
to a slip could imply that, when sufficiently sheared, they are 
sensitive to changes in the stress distribution, and thus easily 
triggered.
From Fig. \ref{fig:contacts}(d) we see that 
more than 80\% of the contacts move more than 2\% of $d_c$ during a 
slip event, and that 15\% move at least 1\% of $d_c$ immediately before 
these slips. 
In some cases, although not consistently, 15\% of the contacts 
move at least 1\% of $d_c$ in connection to micro-slips.
Also, the activity of micro-slips increases during $t_d$, which
suggests that the system becomes more critical.

The time scales of the system spans a factor 100, see Table 
\ref{tab:param}, ranging from 0.16 h to 13.52 h. 
It is tempting to speculate that these time scales 
reflect the spatial dimensions in the system, from 1 mm (diameter
of small contact area) to 10 cm (filling height).
A direct estimate of the maximum length scale can be made from the 
velocities and the observed time scales.
Assuming that the grain--wall contacts in the sandpaper experiment do 
not slip, the internal flow of velocity $v_d=5.4\cdot 10^{-9}$ m/s with the 
characteristic time $\tau_s$ gives a length scale $l_s=260\, \mu$m.
The corresponding length from the initial exponential decrease of 
$m_a$ in an oscillation is $l_0=\tau_0\cdot v_d$ m/s = 74 $\mu$m,
and from the $t_d$ and $t_i$, we get $l_d=t_d\cdot v_d= 101\,\mu$m
and $l_i=t_i\cdot v_i=$115--200$\,\mu$m, respectively.
$v_i$ is the velocity of the bulk during a slip.
The range of $l_i$ results from the different compaction velocities 
found during $t_i$ in the two experiments presented in Fig. \ref{fig:strains}.
The length scales extracted from the characteristic times span 
a smaller range than the time scales do, and are much smaller than the
macroscopic lengths mentioned above.
The small length scales suggest that details of the contact motion
might be of importance to the time scales observed in the system. 

Flow of viscous fluid along a wall can be described by a Navier length 
\cite{pap:deGennes02}.
An average contact velocity, $v_c$, during a slip can be found from knowing 
that contacts slip 6\% of the average contact diameter in 20 minutes,
$v_c=1.15\cdot 10^{-7}$ m/s.
The amount of fluid slip along a wall is given by the Navier length,
$b=\mu/k$, where $mu$ is the fluid viscosity and $k$ is the surface
friction coefficient given by $\sigma/v_c$.
The average shear, $\sigma$, of a contact was found to be between 0.5--1.2 kPa,
thus $k$ is within the range (2.7--11)$\cdot 10^{9}$ Pa\,s/m.
The Navier length is then $b\in[27$--90$] \mu$m, slightly smaller, 
but of the same order as some of the lengths estimated above.

The motion of a contact was not studied with sufficient temporal 
or spatial resolution to conclude whether the whole contact slid a
fraction of $d_c$ at constant velocity, or it slid by self-healing slip pulses 
\cite{pap:Gerde01, pap:Baumberger02}.
Both processes are known from experiments on 
frictional motion in low velocity sheared systems of 
(hard) granular systems \cite{pap:Nasuno98} and slipping of a gel/glass 
interface \cite{pap:Baumberger02}.

\section{Conclusions}
We observe semi-regular oscillations in the measured apparent mass, $m_a$,
at the bottom of a self-compacting ductile grain packing.
The oscillations in one particular experiment are on the order of 
10\% of the total mass $m$ of the grains, and have periods of roughly 6 hours.
The oscillations persist when the granular setup is exchanged with 
a bulk sample of the same ductile material, but disappear when the 
grain--wall interaction is reduced or increased.
Grain--wall contacts are seen to move collectively in correspondence 
to the slip events in $m_a$, as at least 80\% of the contacts move a 
distance larger than 2\% of the average contact diameter during a slip,
see Figs. \ref{fig:Ydisp} and \ref{fig:contacts}.

The decrease of the apparent mass in an oscillation is
thought to be the result of shearing of static wall contacts between
grains and the container wall.
The slow ductile flow internally in the cylinder causes a dynamic stress 
distribution, which results in a continuous increase of 
the shear stress at the grain--wall contacts.
This continuous increase is the reason for the decreasing apparent 
mass.

``Micro-slips'' of the order of 2\% are seen in the normalized apparent 
mass $m_a/m$ during the 
decrease, which probably result from internal stress redistribution in 
granular setups, as they were not seen in $m_a$ of the bulk experiment.
The micro-slips correspond in some cases to limited grain--wall contact
motion, and their `activity' increases during the interval of decreasing
$m_a$.
These slips are also seen when the grain--wall interaction is reduced
or enhanced, that is, when contact motion is stimulated or repressed.

Different characteristic times have been found from curve fitting
of the apparent mass evolution during the `sandpaper' experiment and 
the decreasing part of oscillations in one experiment.
We have also estimated a typical timescale of relaxation of elastic
compressive stresses, and concluded that elasticity is not the driving
mechanism for the observed oscillations.
The characteristic times, together with the period and intervals of
increasing and decreasing $m_a$, are presented in Table \ref{tab:param}.
A successful model should reproduce these characteristic times.

Some attempts at constructing a minimum model have been pursued, but 
the models were discarded as they depended on a finite acceleration
or on unknown microscopic parameters of the system.
Further work is necessary to understand the dynamic behavior of the 
system, and ideas and modeling approaches are welcome.

\section{Acknowledgments}
We wish to thank the Norwegian Research Council for financial support 
through grant number 146031. 
Many thanks also to Thomas Walmann for help with image analysis,
and Renaud Toussaint, Jean Christophe Geminard, Espen Jettestuen, and
Yuri Podladchikov for helpful discussions. Thanks to Bo Nystr{\"o}m
and Anna-Lena Kj{\o}niksen for viscosity measurements.

\begin{thebibliography}{25}
\bibitem[{mis({\natexlab{a}})}]{misc:ma}
\bibinfo{note}{The weight is measured as a force $f$, but the logged data are
  given in units of mass using the relation $m=f/g$, where $g$ is the
  gravitational acceleration.}
\bibitem[{mis({\natexlab{b}})}]{misc:Mettler}
\bibinfo{note}{The vertical change in position was less than a micrometer for
  the loads applied in our experiment. Reference: S. de Villiers, personal
  communication}.
\bibitem[{mis({\natexlab{d}})}]{misc:Unwrap}
\bibinfo{note}{The imaged projection of the curved surface was scaled to that
  of a flat plane, reproducing the surface as seen normal to the curved
  container wall.}
\end{thebibliography}

\end{document}